\documentstyle[twocolumn,prl,aps]{revtex}
\begin{document}
\draft
\title{Excitation Spectrum of Composite Fermions}
\author{X.G. Wu and J.K. Jain}
\address{Department of Physics, State University of New York
at Stony Brook, Stony Brook, New York 11794-3800}
\date{March 15, 1994}
\maketitle
\begin{abstract}

We show that the excitation spectrum of interacting electrons at filling
factor $\nu=\nu^*/(2\nu^*+1)$ is well described in terms of
non-interacting composite fermions at filling factor $\nu^*$, but
does not have a one-to-one correspondence with the excitation spectrum
of non-interacting electrons at $\nu^*$. In particular,
the collective modes of the fractional quantum Hall
states are not analogous to those of the integer quantum Hall states.
We also speculate on the nature of the compressible state at $\nu=1/2$.

\end{abstract}


A composite fermion \cite {Jain} is an
electron carrying two (in general, an even number of) vortices of the
many-body wave function. The wave functions for a many body system of
non-interacting composite fermions at (effective) filling factor
$\nu^*$, $\overline{\Phi}_{\nu^*} ^{n,\alpha}$,
are obtained from the wave functions of non-interacting electrons at
$\nu^*$, $\Phi_{\nu^*}^{n,\alpha}$, as
\begin{equation}
\overline{\Phi}_{\nu^*} ^{n,\alpha} =
{\cal P}D\;\Phi_{\nu^*} ^{n,\alpha}\;\;.
\end{equation}
Here, the Jastrow factor $D=\prod_{j<k}(z_{j}-z_{k})^2$,
where $z_{j}=x_{j}+iy_{j}$ denotes the coordinates of the $j$th
electron, simply attaches two
vortices to each electron of $\Phi_{\nu^*}^{n,\alpha}$ to convert it
into a composite fermion. The operator ${\cal P}$ projects the
state on to the lowest Landau level (LL),
as is appropriate in the large magnetic field
limit. Due to the LL structure, the total (kinetic)
energy of non-interacting electrons at $\nu^*$ is quantized at
$n\hbar\omega_{c}$, measured relative to the ground state energy,
where $\hbar\omega_{c}$ is the cyclotron energy.
The integer $\alpha$ labels the various degenerate many-body states in the
$n$th `band'. The LL's of electrons at $\nu^*$ are mapped into
`quasi-LL's' of composite fermions \cite {quasi}.

According to the composite fermion theory \cite {Jain}, the
strongly correlated liquid of interacting electrons at filling
factor $\nu$ resembles a weakly interacting gas of composite fermions
at filling factor $\nu^*=\nu(1-2\nu)^{-1}$ \cite {Dev,Wu}.
It has been shown in the past \cite {Dev} that the spectrum of interacting
electrons at $\nu$ contains a clearly defined low-energy band
which can be accurately represented as the lowest-energy band of composite
fermions (i.e., by the states $\overline{\Phi}_{\nu^*}^{0,\alpha}$).
In particular, at $\nu=p/(2p + 1)$, where $p$ is an integer, the
ground state contains $p$ filled quasi-LL's of composite fermions,
which explains the fractional quantum Hall effect (FQHE) of
electrons as the integer QHE (IQHE) of composite fermions.

The transformation from electrons to composite fermions in Eq.(1)
has the unusual feature that
the Hilbert spaces of electrons and composite fermions are of
different sizes. For electrons at $\nu^*$, since all higher LL's are
available, there are an infinite number of states.
On the other hand, the Hilbert space of composite fermions at $\nu^*$
is of the same size as the Hilbert space of
electrons at $\nu$ {\em restricted to the lowest LL}, which
is finite (for a finite system). Therefore, any  one-to-one
correspondence between the electron and composite fermions systems
must break down at sufficiently large energies.
This work demonstrates that this happens
right beyond their lowest band. However, the higher bands of
composite fermions continue to provide a good description of
higher energy eigenstates of interacting electrons at $\nu$.
The lack of a one-to-one matching between the
excitation spectrum of interacting electrons at $\nu$ and
non-interacting electrons at $\nu^*$
is of relevance to two issues of current interest.
The first concerns the collective modes of the FQHE states: we find
that many different collective modes of the
IQHE state map into the same collective mode of
the FQHE state. Second, we speculate that the
compressible state at $\nu=1/2$ has the intriguing property that it
possesses a well defined Fermi surface, but it is {\em not} a
regular Landau Fermi liquid.

Our numerical calculations employ the spherical geometry \cite {Haldane}
in which  $N$ electrons move on the surface of a sphere under
the influence of a radial magnetic field produced by a magnetic
monopole of suitable strength placed at the center.
The flux through the surface of the sphere is $N_{\phi}hc/e$, where
$N_{\phi}$ is an integer. We consider spinless electrons confined to
their lowest LL, as appropriate in the limit of large magnetic
field. The eigenstates are labeled by their total orbital angular
momentum $L$, which is analogous to the wave vector of
the planar geometry; larger $L$ corresponds to larger wave vector
\cite {Haldane}.  The Eq.(1) can be easily generalized to spherical
geometry: the Jastrow factor $D$ is now the square of the wave function
of the lowest filled LL, and the interacting electron system at
$N_{\phi}$ is related to the non-interacting composite
fermion system at $N_{\phi}^*=N_{\phi}-2(N-1)$ \cite {Dev}.
Due to the symmetry of the problem, it is sufficient to work in
the sector in which the z-component of the angular momentum
$L_{z}=0$, with the understanding that each state in this
sector represents a degenerate multiplet of $2L+1$ states.

We consider in detail the spectrum of seven electrons
at $N_{\phi}=18$. The low-energy part of the exact Coulomb spectrum
of this  system is shown in Fig.1.
This system corresponds to non-interacting composite
fermions at $N_{\phi}^*=6$. The lowest composite fermion
band contains only one state, with the lowest filled quasi-LL. The wave
function of this state is identical to the Laughlin wave function
for the 1/3 ground state \cite {Laughlin}, which has been tested in detail in
the past \cite {Fano}. The second band of composite fermions
is obtained by exciting a single composite fermion to the next quasi-LL,
which is related to the second band of electrons at $\nu^*=1$.
The latter has states at $L=1,2, ... 7$, but, as found in the
past \cite {Dev}, the $L=1$ state does not produce any composite
fermion state. The other composite fermion states have a good
overlap with the exact eigenstates of Fig.1, as shown in Table II.
The disappearance of the $L=1$ state is the first indication that the
non-interacting electrons and composite fermions have different
excitation spectra.

Next we consider the third band of composite fermions. This is
related to the third ($2\hbar\omega_{c}$) band of non-interacting
electrons at $\nu^*=1$, in which either one
electron has been excited across two LL's, leaving a hole in the
lowest LL, or two electrons have been excited by one LL each, leaving
two holes in the lowest LL. Table I  gives the number of independent
{\em electron} states in this band for all $L$, denoted by $N_{2}(L)$.
We construct the corresponding composite fermion states according
to Eq.(1), and find, surprisingly, that they are not all
linearly independent. They are also not orthogonal to
the composite fermion states of the lower two bands \cite {F4}.
Table I also gives the number, $N_{2}^*(L)$, of the {\em new} (i.e.,
orthogonal to the states of the lower two bands) {\em linearly
independent} composite fermion states in the third band.
Clearly, the third band of composite
fermions contains significantly fewer states than the third band of
non-interacting electrons. In the present example, the total number
of states in the third
band reduces from 75 to 51 upon the composite fermion transformation.
These results dramatically underscore the lack of a one-to-one
correspondence between states in higher bands of non-interacting
composite fermions and non-interacting electrons.

Encouragingly, the $N_{2}^*(L)$ states above
the second band in Fig.1 seem to form a more or less well defined
band. (This would not be true if we took $N_{2}(L)$
states instead, as the analogy to non-interacting {\em electrons}
would suggest.) Furthermore,
the $N_{2}^*$ linearly independent composite fermion states
have significant overlap {\em only} with states in this band, but not
with states outside this band \cite {F8}  -- i.e., they provide a
reasonable basis for the interacting electron states in this band.
For each $L$, we have constructed an {\em orthogonal}
basis of the composite fermion states which is expected to be
close to the exact eigenstates in this band \cite{method}.
The overlaps between the composite fermion states in this basis and
the exact electron states of the second band are shown in Table II.
These are reasonably high, and indicate the validity of the composite
fermion description for higher bands.  (The relatively
poor overlaps for the highest energy states of this band
are to be expected, since they
mix most strongly with states in the higher bands).  We believe
that a similar construction of higher and higher composite fermion
bands will lead to a systematic description of higher and higher
energy eigenstates of Fig.1.

Barring rare coincidences, one does not expect seemingly different
trial wave functions to be {\em mathematically} linearly dependent
(unless, of course, their number exceeds the size of the Hilbert
space). The frequent linear dependence of the composite fermion
trial wave functions provides an indication that they possess certain
non-trivial mathematical symmetries. An appreciation of the formal
structure of the composite fermion states will require a better
understanding of the projection operator.

One might worry that the reduction in the number of states in such
low-energy bands in going from non-interacting electrons to
non-interacting composite fermions
might be a finite size effect.  We do not believe that this is the
case, since, for any given $L$,  the number of states in the first
three bands is only  a small fraction of the total number of states even
for $N=7$ (Table I).

We believe that the qualitative features discovered above in the
context of $\nu=1/3$ are true in general; i.e., the system of
interacting electrons at $\nu=\nu^*/(2\nu^*+1)$ is
well described as a system of  non-interacting composite
fermions at $\nu^*$, but, except for the lowest band,
does not have any simple relation with non-interacting {\em
electrons} at $\nu^*$.

{\em Collective modes}:
The branch in the second band of Fig.1 is interpreted as a
collective mode at small $L$ (i.e., small wave vectors),
as a roton mode at intermediate $L$, and as a
quasiparticle-quasihole excitation at large $L$.
Soon after Laughlin's theory of the ground state at $\nu=1/(2p+1)$,
Girvin, MacDonald and Platzman (GMP) \cite {GMP} developed a single mode
approximation, which  provides an excellent description of
this branch at small and intermediate $L$.
Recently, a Chern-Simons (CS) field theoretical formulation of the
composite fermion theory has been used by Simon and Halperin \cite
{Simon} and by Lopez and Fradkin \cite {Lopez} to investigate the
collective modes of the general $\nu=p/(2p+1)$
FQHE states. These studies predict a {\em series} of
{\em intra}-Landau-level collective modes at $\nu$,
analogous to the inter-LL collective
modes of electrons at $\nu^*=p$ \cite {Kallin}. In our
language, the intra-LL  collective modes of electrons at $\nu$ are
equivalent to
the inter-quasi-LL collective modes of composite fermions at $\nu^*$.
In analogy to the inter-LL electron modes, their wave functions are given by
states in which a single composite fermion is excited from one
of the filled quasi-LL's to an empty quasi-LL.

We consider the issue of the number of collective modes for the 1/3
state. This question is of experimental interest, since the observation
of a  collective mode of the 1/3 state has recently been reported \cite
{Pinczuk}. To this end, we denote states, in which one composite
fermion is excited from the 0th quasi-LL to the $q$th quasi-LL, by
\begin{equation}
\overline{\psi}_{L}(0\rightarrow q)={\cal P}D{\psi}_{L}(0\rightarrow q)
\;\;,
\end{equation}
where ${\psi}_{L}(0\rightarrow q)$ is wave function of electrons (at
$\nu^*=1$) in which one electron has been excited to the $q$th LL,
leaving a hole in the 0th LL.  We have already seen that
the states ${\psi}_{L}(0\rightarrow 1)$ give
a good description of the {\em entire}
low-energy branch in the second band of Fig.1. We have also studied the
$0\rightarrow 2$, $0\rightarrow 3$, and
$0\rightarrow 4$ excitations. We find the surprising
result that
\begin{equation}
\overline{\psi}_{2}(0\rightarrow 2)=\overline{\psi}_{2}(0\rightarrow 1)
\end{equation}
and
\begin{equation}
\overline{\psi}_{3}(0\rightarrow 3)=
\overline{\psi}_{3}(0\rightarrow 2)=\overline{\psi}_{3}(0\rightarrow 1)\;.
\end{equation}
Thus, all composite fermion collective mode states with
$L=2$ and 3 are {\em mathematically} identical \cite {F1}.
The exact equality of these states also holds for
the $\nu=1/3$ systems of 4, 5, and 6 electrons,
and, we believe, is true for arbitrary $N$.
Note that $N_{tot}(L)$ is fairly large for
$L=2$ and 3, and therefore the fact that
the various single-composite-fermion-excitation states are
identical is rather non-trivial. For $L=4$ and 5, the various
composite fermion modes are not identical, but they still have a
reasonably good overlap with $\overline{\psi}_{L}(0\rightarrow 1)$.
These results show that for small $L$, i.e., for small
wave vectors, different {\em electron} collective modes at
$\nu^*=1$ produce a single collective mode at $\nu=1/3$ \cite {F10}.
This again emphasizes that the {\em excitations} of the FQHE system at $\nu$
do not resemble the excitations of the non-interacting
{\em electron} system at $\nu^*$.

We have also carried out a preliminary study of the collective modes
of other FQHE states. Here, the situation is more
complicated, since several quasi-LL's are occupied. Our findings
are as follows. (i) The low-energy branch (in the second band)
exhausts (to a good approximation) the spectral weight of the GMP
collective mode excitation {\em at small} $L$, suggesting only one
intra-LL collective mode at small wave vectors. (ii)
$\overline{\psi}_{L}(p-1\rightarrow p)$ always
provides a good description of the entire low-energy branch.
(iii) For non-interacting  electrons, there are in general several
degenerate inter-LL collective modes, which may hybridize.
For 2/5, one linear combination of $\overline{\psi}_{L}(1\rightarrow
3)$ and $\overline{\psi}_{L}(0\rightarrow 2)$
is almost identical to the
$\overline{\psi}_{L}(1\rightarrow 2)$ for small $L$, and
does not produce any new collective mode. The orthogonal
combination is not close to any single electron eigenstate.
Further work is in progress.

It might be argued that our conclusions might not remain true in
the presence
of a small amount of LL mixing, always present in experiment. In this
case, it would seem natural {\em not} to project the composite fermion
states on to the lowest LL, so they would not be identical.
However, clearly, the unprojected composite fermion states
will simply provide {\em different approximations} to the {\em same}
collective mode. The essential point, quite obvious on physical
grounds, is that the {\em intra-LL} spectrum of interacting
electrons, or the {\em number} of {\em intra}-LL collective modes,
cannot change in any essential manner when a small amount of LL
mixing is allowed. Therefore, our conclusions, obtained with the
lowest LL approximation, should remain valid for the more
realistic situation of a large though not infinite $B$.

{\em $\nu=1/2$ state}:
Before closing, we comment on the possible implications of
our results on another issue of current interest, namely
on the nature of the compressible state at $\nu=1/2$.
Consider the
$p/(2p+1)$ sequence of incompressible FQHE states. At low
temperatures, the thermodynamic properties of the incompressible FQHE
states are
dominated by the excitation gap, and the above considerations, which
concern the spectrum {\em above} the gap, are not of much relevance.
However, the gap vanishes in the limit $p\rightarrow\infty$,
as the sequence approaches $\nu=1/2$,
and the excited states become relevant
to the thermodynamics.  Two comments
pertaining to this limit may be made, {\em provided}
it is assumed that  the composite fermion description remains valid
for arbitrarily large $p$ (which is unfortunately not testable in
finite system studies \cite {Rezayi}).
(i) The {\em ground state} of interacting electrons at $\nu=1/2$
resembles a Fermi sea of composite fermions with a well defined Fermi
surface, as proposed by Halperin, Lee, and Read \cite {HLR}. Recent
experiments lend strong support to the existence of a Fermi sea at
$\nu=1/2$ \cite {cfex}.  (ii) However, the {\em excitations} of the
1/2 system do {\em not} have a one-to-one correspondence
with the excitations of {\em electrons} at zero magnetic field.
Therefore, despite the existence of the Fermi
surface, the $\nu=1/2$ state is not a regular Landau Fermi
liquid (it may be termed ``composite fermion liquid"). Nevertheless,
since it {\em is related} to the electron Fermi liquid, a Fermi-liquid-like
description with renormalized parameters might be valid.

In conclusion, while the composite fermion trial wave functions
provide a good description of the excitation spectrum of interacting
electrons in the FQHE regime, a
mean-field treatment, which implies a one-to-one correspondence
between the states of interacting electrons at $\nu=\nu^*/(2\nu^*+1)$
and non-interacting electrons at $\nu^*$, is not valid beyond the
lowest band.  This work was supported in part by NSF under Grant No.
DMR93-18739.

Fig.1 Some low-energy lowest-LL eigenstates of seven electrons at
$N_{\phi}=18$. For full spectrum, see \cite {Haldane}.

\vspace{3cm}

\begin{center}

\begin{tabular}{|c|c|c|c|c|} \hline
$L$& $N_{0}$ ($N_{0}^*$) & $N_{1}$ ($N_{1}^*$)
& $N_{2}$ ($N_{2}^*$)  & $N_{tot}$ \\ \hline
0  & 1(1)   & 0 (0)  & 3 (2)  & 10 \\ \hline
1  & 0(0)   & 1 (0)  & 3 (2)  & 12 \\ \hline
2  & 0(0)   & 1 (1)  & 9 (5)  & 29 \\ \hline
3  & 0(0)   & 1 (1)  & 8 (4)  & 33 \\ \hline
4  & 0(0)   & 1 (1)  & 11 (7)  & 48 \\ \hline
5  & 0(0)   & 1 (1)  & 9 (5)  & 49 \\ \hline
6  & 0(0)   & 1 (1)  & 10 (7)  & 65 \\ \hline
7  & 0(0)   & 1 (1)  & 7 (5)  & 64 \\ \hline
8  & 0(0)   & 0 (0)  & 7 (6)  & 75 \\ \hline
9  & 0(0)   & 0 (0)  & 3 (3)  & 74 \\ \hline
10  & 0(0)   & 0 (0)  & 3 (3)  & 83 \\ \hline
11  & 0(0)   & 0 (0)  & 1 (1)  & 78 \\ \hline
12  & 0(0)   & 0 (0)  & 1 (1)  & 86 \\ \hline
\end{tabular}

\end{center}

Table I. The number of independent many-body
electron and composite fermion states in the $n$th band, denoted by
$N_{n}(L)$ and $N_{n}^*(L)$, respectively, for the
first three bands $n=0,1,2$. $N_{tot}(L)$ is the total number of states
at $L$ for the system of Fig.1.

\begin{center}

\begin{tabular}{|l|l|} \hline
$L$& overlap \\ \hline
0  & 0.9929, 0.7909 \\ \hline
1  & 0.9977, 0.8506 \\ \hline
2 & (0.9031), 0.9531, 0.9550, 0.9819, \\
  & 0.9224, 0.7993 \\ \hline
3 & (0.9793), 0.9793, 0.9909, 0.9543, \\
  & 0.9527 \\ \hline
4 & (0.9970), 0.9949, 0.9970, 0.9941, \\
  & 0.9784, 0.9516, 0.9962, 0.7901 \\ \hline
5 & (0.9906), 0.9925, 0.9796, 0.9788, \\
  &  0.9486, 0.9229 \\ \hline
6 & (0.9822), 0.9914, 0.9841, 0.9861, \\
  &  0.9887, 0.9685, 0.9516, 0.8735 \\ \hline
7 & (0.9853), 0.9842, 0.9872, 0.9798, \\
  & 0.9287, 0.9070 \\ \hline
8 & 0.9943, 0.9931, 0.9921, 0.9867 \\
  & 0.9839, 0.8735 \\ \hline
9 & 0.9911, 0.9947, 0.9684 \\ \hline
10 & 0.9947, 0.9915, 0.9702 \\ \hline
11 & 0.9876 \\ \hline
12 & 0.9765 \\ \hline
\end{tabular}

\end{center}

Table II. The overlaps of a suitably chosen
orthogonal linear combination of composite fermion states in the second
(in parentheses) and third bands with the corresponding exact states
of Fig.1 (in order of increasing energy).


\begin{thebibliography}{99}
\bibitem{F4} A curious feature is that the third band contains the
states of the lower two bands; i.e., the composite fermion states of
the lower bands can be exactly expressed as linear combinations of
the third band states.
\bibitem{F8} Remember that they are already orthogonal to the states of
the lower bands.
\bibitem{method} We construct this composite fermion basis as follows.
Call the initial composite fermion states
$\phi_{1}, ..., \phi_{M}$ and the exact Coulomb eigenstates in the
third band of Fig.1 $\psi_{1}, ..., \psi_{M}$. We construct $\phi'_{i}
=\sum_{j} c_{ij}\phi_{j}$, where $c_{ij}=<\psi_{i}|\phi_{j}>$. If the
two bases were related by an orthogonal transformation, then we would
have $\phi'_{i}=\psi_{i}$. This, however, is not the case,
and $\phi'_{i}$ are only approximately orthogonal. We
Gramm-Schmid orthogonalize them to obtain $\phi''_{i}$
used to calculate the overlaps in Table II.
\bibitem{F1} These are the only collective mode states at $L=2$
and $L=3$.
\end{thebibliography}
\end{document}